\documentclass[nofootinbib,preprint,12pt,aps]{revtex4}%
\usepackage{amsmath}
\usepackage{graphicx}
\usepackage{version}
\usepackage{amsfonts}
\usepackage{amssymb}%
\providecommand{\U}[1]{\protect \rule{.1in}{.1in}}

\begin{document}
\title{Neutrino mixing and discrete symmetries}
\author{Bo Hu}
\email{bohu@ncu.edu.cn}
\affiliation{Department of Physics, Nanchang University, Jiangxi 330031, China}

\begin{abstract}
In this paper we discuss a new way to derive neutrino mixing patterns,
which\ originates from the idea proposed in a recent article by Hernandez and
Smirnov. Its applications to various cases are discussed. We first present the
complete set of possible mixing patterns for the minimal case where unbroken
residual symmetries of the Majorana neutrino and left-handed charged-lepton
mass matrices obey some general assumptions that are also satisfied by many
models based on discrete symmetries. We find that they are either well-known
mixing patterns or phenomenologically disfavored ones. It shows clearly that,
for full-mixing matrices to fit the mixing data with small or negligible
corrections, it is necessary to go beyond the minimal scenario. We present an
explicit formalism for a rather general nonminimal case. Some applications and
phenomenological implications are discussed. Several new mixing patterns are derived.

\end{abstract}
\maketitle

\section{Introduction}

The discovery of neutrino oscillations provides clear evidence for physics
beyond the Standard Model (see, e.g.~\cite{Altarelli1, Sim3} and for recent
global fits, see \cite{Tortola, Fogli, Gonzalez}). In this work we study
neutrino mixing determined by underlying discrete flavor symmetries \cite{dfr,
dfr1}. Models based on this approach have been discussed extensively (see,
e.g., \cite{dm18, dm19, dm1, dm2, dm3, dm4, dm5, dm22, dm23, dm6, dm7, dm20,
dm8, dm21, dm9, dm10, dm17, dm11, dm12, dm13, dm14, dm24, dm15, dm16} for some
recent works) and they often lead to intriguing mixing patterns, including the
well-known tribimaximal mixing (TBM) \cite{tbm} and bimaximal mixing (BM)
\cite{bm}. However, a challenge is posed by recent experimental results on the
reactor mixing angle \cite{T2K, MINOS, CHOOZ, RENO, DAYA-BAY} because many
models discussed previously did not predict the measured value of $\theta
_{13}$ (see, e.g. \cite{Antusch} or \cite{Albright} for a survey of model
predictions). Although good agreement with the data can be achieved by
introducing corrections or other means (see, e.g. \cite{HHJ2, HHJ1, xing1,
xing3}), it is still worthwhile to seek for symmetries and mixing patterns
that are in good agreement with the data without introducing large corrections
which may badly break the (would-be) residual symmetries, e.g. the THF mixing
or bi-trimaximal mixing discussed in recent articles \cite{toorop, ding1,
king}.

In this paper we focus on Majorana neutrinos. The mass matrix of Majorana
neutrinos has a $\mathbf{Z}_{2}\times \mathbf{Z}_{2}$ symmetry, and the
left-handed charged lepton mass matrix has a $U(1)\times U(1)$ symmetry
\cite{Grimus, lam3} if they are required to belong to $SU(3)$. When restricted
to discrete symmetries, the latter is usually reduced to a $\mathbf{Z}_{m}$ or
other finite symmetry belonging to the $U(1)\times U(1)$ (for some examples,
see \cite{toorop}). In models based on discrete symmetries, these symmetries
may coincide with the symmetries preserved by relevant mass terms after the
breaking of flavor symmetry,\footnote{Models with completely broken symmetries
can be constructed (e.g., \cite{ding3}), which are beyond the scope of this
work and hence will not be considered here.} which is the case that will be
considered in this paper. We follow \cite{Sim1} and carry out a model
independent study of neutrino mixing. The basic assumption we adopt is that
the residual symmetries of the mass matrices belong to an underlying discrete
flavor symmetry. Our main concern is the general results that can be drawn
from this assumption.

We find that the idea of \cite{Sim1} leads to a new way to derive mixing
matrices, which is simpler than the group theoretical method. It does not
require a flavor group to be specified in advance. On the contrary, it can
provide the necessary conditions from which one can find the flavor symmetry
corresponding to a mixing matrix or mixing-matrix column derived by this
method. If the flavor symmetry is known, it can serve as a quick way to derive
mixing matrices or elements and cross-check with group theoretical
calculations. From the equation constraining the residual symmetries, an
interesting result we find is that, under some general and plausible
assumptions which are satisfied by many models and adopted in \cite{Sim1}, one
can prove rigorously that phenomenologically viable mixing patterns that
discrete symmetries can lead to include only TBM, BM and the golden ratio
mixings. It also provides a different way to approach some known results. For
example, one can show that the minimal horizontal symmetry that can produce
TBM as a full-mixing matrix is $\mathbf{S}_{4}$, which was first pointed out
in \cite{lam2}.

As suggested by the no-go result mentioned above, those assumptions may have
to be relaxed, and hence new mixing patterns can be obtained. Explicit
formalism for a rather general nonminimal case is also provided in this paper.
Several interesting applications are discussed and new mixing matrices are
presented. Their phenomenological implications are also discussed briefly. We
find that, although not impossible at all, it is still difficult to find a
mixing matrix that can produce all the mixing data. This result agrees with
\cite{Lam1} and \cite{Hol1} in which the results of the searches for groups up
to certain large orders are reported.\ 

Although our approach is based on that of \cite{Sim1}, the concerns and
discussions differ quite a lot. We also note that in a recent work
\cite{Sim2}, authors of \cite{Sim1} also generalize the results of their
previous work. But the subjects and the formalism are still quite different.
More attention is paid to partial mixings and phenomenology in \cite{Sim2},
and their discussions are more group theoretical oriented. Most of our
attention is concentrated on full-mixing patterns. A general result about the
minimal scenario of \cite{Sim1} is presented and discussed in this paper.
Besides the formalism that can be used in general cases, we also provide a
simple formalism for a rather general nonminimal scenario from which most
known mixing patterns, including those discussed in \cite{Sim2} or
\cite{toorop}, can be derived. It also provides a different way to understand
the permutation feature of some mixing patterns. Several new mixing patterns
are also derived and presented.

The paper is organized as follows. In Sec.~II, we review briefly the idea
proposed in \cite{Sim1}. For later convenience, the formalism for squared
mixing elements, which can be used in general cases, is presented in this
section. In Sec.~III, the attention is focused on the minimal scenario. A
general result concerning possible mixing patterns is presented and discussed.
Then, in Sec.~IV, the consequences of relaxing some of the assumptions adopted
in the minimal scenario and the relevant phenomenology are discussed. We
summarize in Sec.~V.

\section{Neutrino mixing and discrete symmetries}

We first set up our notations. We denote the generator of the symmetry of the
left-handed charged-lepton mass matrix (i.e., $M_{l}^{\dag}M_{l}$ where
$M_{l}$ is the mass matrix of charged leptons) by $T$ and those of the
Majorana neutrino mass matrix by $S_{i}$ where $i=1$, $2$, or $3$. The flavor
symmetry is denoted by $G_{f}$, and the symmetries generated by $T$ and
$S_{i}$ are denoted by $G_{e}$ and $G_{\nu}$, respectively. In this work, we
always assume that $G_{\nu}=\mathbf{Z}_{2}\times \mathbf{Z}_{2}$ and $T$
generates a $\mathbf{Z}_{m}$ symmetry. We also assume that $T$ and $S_{i}$
belong to $SU(3)$ and, for simplicity, $\mathrm{Tr}[T]$ is real in most of our
discussions. Some comments on these assumptions are given at the end of this section.

In the basis where the charged lepton mass matrix is diagonal, $T$ can be
written as a diagonal matrix denoted by $T_{d}$. In the case where
$\mathrm{Tr}[T]$ is real, $T_{d}$ can be written as one of the following three
matrices:%
\begin{align}
T_{e} &  \equiv T_{1}=\mathrm{diag}\{1,e^{2\pi ik/m},e^{-2\pi ik/m}%
\},\nonumber \\
T_{\mu} &  \equiv T_{2}=\mathrm{diag}\{e^{2\pi ik/m},1,e^{-2\pi ik/m}%
\},\label{ti}\\
T_{\tau} &  \equiv T_{3}=\mathrm{diag}\{e^{2\pi ik/m},e^{-2\pi ik/m}%
,1\},\nonumber
\end{align}
since it belongs to $SU(3)$. Therefore,%
\begin{equation}
\mathrm{Tr}[T]=1+2\cos2\pi k/m.\label{trt}%
\end{equation}
when $\mathrm{Tr}[T]$ is real. In addition, we assume that $T_{\alpha}$
($\alpha=e$, $\mu$, or $\tau$) is nondegenerate or $m\geq3$; otherwise charged
lepton masses cannot be guaranteed to be nondegenerate and ambiguities in
mixings may arise \cite{lam3}. Further discussions can be found at the end of
this section.

The key assumption on $T$ and $S_{i}$ is that they belong to $G_{f}$, a finite
flavor symmetry. It then follows that $(S_{i}T)^{p_{i}}=\mathbb{I}$ for some
$p_{i}$ which leads to \cite{Sim1}\footnote{Other relations are possible if
$T$ or $S_{i}$ appear in $G_{f}$ indirectly.}%
\begin{equation}
(S_{i}T)^{p_{i}}=(U_{PMNS}S_{i}^{d}U_{PMNS}^{\dag}T_{\alpha})^{p_{i}%
}=\mathbb{I,}\label{stp}%
\end{equation}
where $U_{PMNS}$ is the lepton mixing matrix and $S_{i}^{d}$ are diagonal
matrices given by%
\begin{equation}
S_{1}^{d}=\mathrm{diag}\{1,-1,-1\},\quad S_{2}^{d}=\mathrm{diag}%
\{-1,1,-1\},\quad S_{3}^{d}=S_{1}^{d}S_{2}^{d}.\label{si}%
\end{equation}

Then, denote $(S_{i}T)^{-1}=T^{m-1}S_{i}$ by $W_{i}$. One finds that
\begin{equation}
\left(  W_{i}\right)  ^{p_{i}}=\left[  (S_{i}T)^{-1}\right]  ^{p_{i}%
}=\mathbb{I}\mathcal{.}\label{wp}%
\end{equation}
Our basic assumption on the residual symmetries can then be summarized by the
relations%
\begin{equation}
S_{i}^{2}=T^{m}=\left(  W_{i}\right)  ^{p_{i}}=S_{i}TW_{i}=\mathbb{I.}%
\label{ba}%
\end{equation}
The characteristic equation of $W_{i}$\ can be written as \cite{Sim1}%
\begin{equation}
\lambda^{3}+w_{i}\lambda^{2}-w_{i}^{\ast}\lambda-1=0,\label{ceq}%
\end{equation}
where $w_{i}=-\mathrm{Tr}[W_{i}]$. Because of Eq.(\ref{wp}), one has
$\lambda^{p_{i}}=1$, and from the equation above, it follows that if $w_{i}$
is real, the eigenvalues of $W_{i}$ are given by $\lambda=1$, $e^{2\pi
in_{i}/p_{i}}$, or $e^{-2\pi in_{i}/p_{i}}$, and hence $w_{i}=-1-2\cos2\pi
n_{i}/p_{i}$.

Until now we have followed \cite{Sim1} in which more details can be found.
Below we shall discuss first the general situation. Later in the next section
we return back to the minimal scenario discussed in \cite{Sim1}. Since
$\mathrm{Tr}[W_{i}]=\mathrm{Tr}[(S_{i}T)^{-1}]=\mathrm{Tr}[(S_{i}T)^{\dag
}]=\left(  \mathrm{Tr}[S_{i}T]\right)  ^{\ast}$, one finds that%
\begin{equation}
w_{i}^{\ast}=-\left(  \mathrm{Tr}[W_{i}]\right)  ^{\ast}=-\mathrm{Tr}%
[U_{PMNS}S_{i}^{d}U_{PMNS}^{\dag}T_{d}]=\mathrm{Tr}[T_{d}]-2\mathrm{Tr}\left[
A_{i}T_{d}\right]  \label{wia}%
\end{equation}
where $A_{i}$ are defined as%
\[
A_{i}=\mathrm{diag}\{|\left(  U_{PMNS}\right)  _{1i}|^{2},|\left(
U_{PMNS}\right)  _{2i}|^{2},|\left(  U_{PMNS}\right)  _{3i}|^{2}\}.
\]
Equation (\ref{wia}) is the most general formalism for mixing elements in this
framework. As will be shown later, it can be written in more convenient and
explicit forms.

If all $S_{i}\in G_{f}$, an important condition can be derived from
Eq.(\ref{wia}). Summing over $i$ leads to%
\begin{equation}
\sum_{i=1}^{3}w_{i}^{\ast}=-\sum_{i=1}^{3}\left(  \mathrm{Tr}[W_{i}]\right)
^{\ast}=\mathrm{Tr}[T_{d}]\label{ucon0}%
\end{equation}
which follows from $\sum_{i=1}^{3}S_{i}^{d}=-\mathbb{I}$ and the unitarity of
$U_{PMNS}$. This equation will be referred to as the unitarity condition in
this paper. Since this condition always holds, it must be obeyed by any
combination of $S_{i}$ and $T$ that generates a full-mixing matrix. As
Eq.(\ref{wia}), it can also be written in more explicit forms. Equations
(\ref{wia}) and (\ref{ucon0}) are the starting point of our discussion that follows.

When $\mathrm{Tr}[T]$ is real, $T_{d}$ is given by one of the $T$ matrices in
Eq.(\ref{ti}). Then from Eq.(\ref{wia}) it follows that
\begin{align}
\operatorname{Re}[w_{i}] &  =1-2|\left(  U_{PMNS}\right)  _{\alpha i}%
|^{2}\left(  1-\cos2\pi k/m\right)  \label{meq2}\\
\operatorname{Im}[w_{i}] &  =2\left[  |\left(  U_{PMNS}\right)  _{\beta
i}|^{2}-|\left(  U_{PMNS}\right)  _{\gamma i}|^{2}\right]  \sin2\pi
k/m\label{meq1}%
\end{align}
where $\beta,\gamma \neq \alpha$ and $\beta<\gamma$. Solutions to
Eqs.(\ref{meq2}) and (\ref{meq1}) should respect the conditions
\begin{equation}
0\leq|\left(  U_{PMNS}\right)  _{\rho i}|^{2}\leq1,\quad \rho=\alpha
,\, \beta,\, \mathrm{or}\, \gamma \label{rc}%
\end{equation}
which will be referred to as reality conditions. Explicit expressions for
$|\left(  U_{PMNS}\right)  _{\rho i}|^{2}$ will be given later. Note that the
value of $\alpha$ depends on which $T$ matrix given in Eq.(\ref{ti}) is used
in Eq.(\ref{wia}). Using a different $T$ matrix but keeping $m$ and $k$ fixed
results in a reordering of the elements in the mixing vector by which we mean
a column of the mixing matrix. One may use any $T$ matrix given in
Eq.(\ref{ti}) as long as the results are consistent with the experimental data.

Together with the unitarity and the reality conditions, the above formalism
provides a simple way to derive mixing matrix or mixing elements. For example,
it was shown in \cite{toorop} that the group $PSL(2,Z_{7})$ can lead to a
mixing matrix given by\footnote{Following \cite{toorop}, we use $\left \Vert
U_{PMNS}\right \Vert $ to denote the matrix with every entry being the absolute
value of the corresponding one in the mixing matrix $U_{PMNS}$.}%
\begin{equation}
\left \Vert U_{PMNS}\right \Vert =\frac{1}{2}\left(
\begin{array}
[c]{ccc}%
\sqrt{\frac{1}{2}(3+\sqrt{7})} & 1 & \sqrt{\frac{1}{2}(3-\sqrt{7})}\\
1 & \sqrt{2} & 1\\
\sqrt{\frac{1}{2}(3-\sqrt{7})} & 1 & \sqrt{\frac{1}{2}(3+\sqrt{7})}%
\end{array}
\right)  .\label{psl2z7u}%
\end{equation}
The representation matrices of $PSL(2,Z_{7})$ are given by%
\begin{equation}
F=\frac{2}{\sqrt{7}}\left(
\begin{array}
[c]{ccc}%
s_{1} & s_{2} & s_{3}\\
s_{2} & -s_{3} & s_{1}\\
s_{3} & s_{1} & s_{2}%
\end{array}
\right)  ,\quad G=\frac{2}{\sqrt{7}}\left(
\begin{array}
[c]{ccc}%
e^{4\pi i/7} & 0 & 0\\
0 & e^{2\pi i/7} & 0\\
0 & 0 & e^{8\pi i/7}%
\end{array}
\right)  \label{psl2z7g}%
\end{equation}
where $s_{k}=\sin k\pi/7$. $S_{1}$, $S_{2}$, and $T$ can be chosen as
$F,G^{2}FG^{3}FG$, and $G^{3}F$, respectively. The $\left \Vert U_{PMNS}%
\right \Vert $ matrix given in Eq.(\ref{psl2z7u}) can be obtained by
diagonalizing $S_{i}$ and $T$. The formalism given above provides a different
but more efficient way to derive it. More explicit formalism and examples will
be given later.

Now we have all the necessary ingredients, but before proceeding, there are a
few comments we would like to make.

\begin{enumerate}
\item Since $T\in SU(3)$, if $\mathrm{Tr}[T]$ is not real, then it does not
have a $+1$ eigenvalue and hence no vacuum alignment can break $G_{f}$ into
the $\mathbf{Z}_{m}$ symmetry generated by $T$. It can be arranged that
residual symmetries are preserved indirectly \cite{ding1, ding3}, but it
sounds more natural if they emerge directly, and thus, except in an example
given in Sec.~IV, we assume in most of our discussions that $\mathrm{Tr}[T]$
is real.

\item When degeneracy occurs among the eigenvalues of $T$, whether
$\mathrm{Tr}[T]$ is real or not, the lepton mixing matrix cannot be determined
unambiguously and the solutions to Eq.(\ref{wia}) are not unique.
Nevertheless, the vanishing of $\operatorname{Im}(\mathrm{Tr}[T])$ assures
that for any $m>2$, the eigenvalues of $T$ are nondegenerate and the mixing
matrix can be determined unambiguously. As to the case where $m=2$, to
eliminate the ambiguity, one may enlarge the $\mathbf{Z}_{2}$ symmetry
generated by $T$ to a larger one, e.g. a $\mathbf{Z}_{2}\times \mathbf{Z}_{2}$.
An example is given in footnote 4. Since we are interested in general cases,
we will not consider this particular case further.

\item Unlike the assumption on $\mathrm{Tr}[T]$, $\mathrm{Tr}[W_{i}]$ are
assumed to be real mostly for calculational simplicity. In many cases,
including the $PSL(2,Z_{7})$ example discussed above, $\mathrm{Tr}[W_{i}]$ are
not real. But we should mention that this assumption is indeed satisfied by
many models leading to well-known mixing patterns (e.g., TBM, BM, and the
golden ratio mixings). In addition, it can be treated as a reasonable
phenomenological assumption. The reason is that the equalities between the
absolute values of mixing elements are still phenomenologically viable, and
from Eq.(\ref{meq1}) it follows that when Eq.(\ref{meq4}) is satisfied,
$\operatorname{Im}[w_{i}]$ vanishes automatically. Detailed discussion about
the case where $\mathrm{Tr}[W_{i}]$ are not real will be presented in Sec.~IV.

\item From experimental data, it follows that in the minimal scenario,
$\sin \theta_{13}$ can only result from Eq.(\ref{meq3}) (see below),\ and
hence, to accommodate measured $\sin \theta_{13}\sim0.15$, $1-w_{i}$ must be
almost vanishing because $\sin^{2}\theta_{13}\sim0.02$, which requires a large
$p_{i}$ (recall that $w_{i}=-1-2\cos2\pi n_{i}/p_{i}$) such that $n_{i}/p_{i}$
can be close to $1/2$. Numerical computation also shows that to fit the data,
$p_{i}$ should be larger than $10$. Large $p_{i}$ corresponds to a $S_{i}T$ of
large order. Therefore, the larger the $p_{i}$ is, the less the chance for
small groups to accommodate the data. Allowing corrections to $\sin \theta
_{13}$ may significantly lower the requirement, but introducing large
corrections may also require the price of weakening the role of symmetry to be
paid. More discussion can be found in Sec.~IV.

\item If all $S_{i}\in G_{f}$ and $m\geq3$, then the matrix $\left \Vert
U_{PMNS}\right \Vert $ is completely determined by $S_{i}$ and $T$ and is
referred to as a full-mixing matrix. In the case where both $\mathrm{Tr}[T]$
and $\mathrm{Tr}[S_{i}T]$ are real, one can show that Eq.(\ref{meq4}) (see
below) always leads to a vanishing mixing element or a maximum Dirac phase in
an appropriate parametrization \cite{xing2}. The vanishing element implies
that $w_{i}=1$, and thus from Eq.(\ref{ceq}), one finds that $p_{i}=2$ and
$(S_{i}T)^{2}=1$, i.e. one $S_{i}T$ must be of order $2$, which can also be
seen from the result presented in the next section. Note that $p_{i}$ cannot
be completely determined in this way.
\end{enumerate}

\section{Mixing patterns in the minimal scenario}

In this section, we concentrate on the full-mixing patterns in the minimal
scenario where both $\mathrm{Tr}[T]$ and $\mathrm{Tr}[W_{i}]$ are real. As
discussed above, the residual symmetries are strongly constrained by the
unitarity condition Eq.(\ref{ucon0}). Thanks to Eq.(\ref{ba}), $\mathrm{Tr}%
[T]$ and $\mathrm{Tr}[W_{i}]$ depend on the orders of $T$ and $W_{i}$, which
can then be solved from Eq.(\ref{ucon0}). Obviously it admits many solutions.
What is interesting and somewhat surprising in the minimal case is that the
complete set of solutions can be determined. Consequently, all the possible
mixing patterns can also be derived. After some general discussions about the
solutions, the one that leads to a golden ratio mixing will be discussed in
detail to demonstrate the way to derive the mixing pattern and the flavor
symmetry corresponding to a particular solution and clarify its difference
from the group theoretical method.

We begin with Eqs.(\ref{meq2}) and (\ref{meq1}) which determine the mixing
elements. When both $\mathrm{Tr}[W_{i}]$ and $\mathrm{Tr}[T]$ are real, they
can be written as \cite{Sim1}%
\begin{align}
|\left(  U_{PMNS}\right)  _{\alpha i}|^{2} &  =\frac{1-w_{i}}{4\sin^{2}%
\frac{k\pi}{m}}=\frac{1+\mathrm{Tr}[W_{i}]}{4\sin^{2}\frac{k\pi}{m}%
}\label{meq3}\\
|\left(  U_{PMNS}\right)  _{\beta i}|^{2} &  =|\left(  U_{PMNS}\right)
_{\gamma i}|^{2}=\frac{1}{2}\left(  1-|\left(  U_{PMNS}\right)  _{\alpha
i}|^{2}\right)  .\label{meq4}%
\end{align}

We also assume that all $S_{i}\in G_{f}$. Hence the unitarity condition given
by Eq.(\ref{ucon0}) must be respected. As discussed below Eq.(\ref{ceq}), if
$\mathrm{Tr}[W_{i}]$ is real, then
\begin{equation}
\mathrm{Tr}[W_{i}]=1+2\cos \frac{n_{i}}{p_{i}}2\pi \label{twia}%
\end{equation}
where $p_{i}$ is the order of $W_{i}$. Note that the greatest common divisor
of $n_{i}$ and $p_{i}$, $\gcd(n_{i},p_{i})=1$.\ Now from Eqs.(\ref{trt}) and
(\ref{ucon0}), one has%
\begin{equation}
1+2\cos \frac{k}{m}2\pi=-\sum \limits_{i=1}^{3}\mathrm{Tr}[W_{i}]=-3-2\sum
\limits_{i=1}^{3}\cos \frac{n_{i}}{p_{i}}2\pi \label{twia1}%
\end{equation}
which can be written as
\begin{equation}
\sum \limits_{j=1}^{4}2\cos \frac{n_{j}}{p_{j}}2\pi=-4\label{ucon}%
\end{equation}
where $k$ is replaced by $p_{4}$ and $m$ by $n_{4}$. Without loss of
generality, we require that $0<n_{j}/p_{j}\leq1/2$. Equation (\ref{ucon}) is a
necessary condition that $p_{j}$, the orders of $T$ and $W_{i}$ (or $S_{i}T$)
must obey. Together with $n_{i}$, they can be used to construct explicitly the
(minimal) group corresponding to a solution. Its relation with the group
theoretical method will be discussed shortly.

We find that a complete and rigorous solution to Eq.(\ref{ucon}) can be
derived by using algebraic number theory \cite{poll, lang}, which can
translate Eq.(\ref{ucon}) into a much simpler arithmetic equation. Although a
rigorous derivation is interesting by its own right, it is somewhat lengthy
because some mathematical concepts and results need to be introduced. In
addition, the solutions given below can be verified numerically by evaluating
Eq.(\ref{ucon}) continuously until $p_{i}$ reach a desired large value.
Therefore, in order to concentrate on physical discussion, we will present the
detailed mathematical derivation elsewhere.

Since $n_{i}$ can be easily found from Eq.(\ref{ucon}) when $p_{i}$ are given,
below we denote the solution to Eq.(\ref{ucon}) by $\{p_{j}\}$. Although
expected, it is still a little bit surprising to find that, besides obviously
disfavored solutions, i.e., $\{1,2,2,2\}$ and $\{2,2,p_{3},p_{4}\}$,
Eq.(\ref{ucon}) admits only three other solutions: $\{3,3,3,3\}$,
$\{2,3,3,4\}$, and $\{2,3,5,5\}$. Note that $p_{3}$ and $p_{4}$ in the second
solution satisfy $n_{3}/p_{3}+n_{4}/p_{4}=1/2$, where $n_{3}$ and $n_{4}$ are
arbitrary integers satisfying $0<n_{j}/p_{j}\leq1/2$ and $\gcd(n_{j},p_{j})=1$.

To find the $\left \Vert U_{PMNS}\right \Vert $ matrix corresponding to a
solution, one must first assign one $p_{j}$ to $m$ and the others to
$p_{1,2,3}$ in Eq.(\ref{twia}). It is obvious that the first solution
$\{1,2,2,2\}$ does not respect the assumption that $m\geq3$, and hence it
cannot lead to an unambiguous mixing matrix. The second solution
$\{2,2,p_{3},p_{4}\}$ is also phenomenologically disfavored because of the
same reason, which requires that any $p_{j}=2$ can only be assigned to the
order of an $S_{i}T$. From Eqs.(\ref{meq3}) and (\ref{twia}) it follows that
$U_{\alpha i}=0$ if $p_{i}=2$. However, $p_{1,2}=2$ results in two vanishing
elements which lead to two vanishing mixing angles and hence this solution is
not phenomenologically favorable.

The third solution $\{p_{i},m\}=\{3,3,3,3\}$ leads to an interesting
$\left \Vert U_{PMNS}\right \Vert $ matrix in which all the elements equal
$1/\sqrt{3}$, but it is also not phenomenologically favorable because of its
large deviation from the experimental data. Note that it leads to a maximum
Dirac CP phase.

Therefore, we are left with the last two solutions. From Eqs.(\ref{meq3}%
)--(\ref{twia}), it is easy to show that the assignment $\{p_{1},p_{2}%
,p_{3},m\}=\{3,3,2,4\}$, $\{4,3,2,3\}$, $\{3,5,2,5\}$, or $\{5,5,2,3\}$ leads
to BM, TBM, or the golden ratio mixings, respectively.\footnote{As discussed
in the last section, we require that $m\neq2$. If $T^{2}=\mathbb{I}$, to fix
the mixings, one may enlarge the $\mathbf{Z}_{2}$ symmetry generated by $T$ to
a larger one. For example, if $\{3,5,5,2\}$ is assigned to $\{p_{i},m\}$ and
$G_{e}$ is chosen to be $\mathbf{Z}_{2}\times \mathbf{Z}_{2}$, then $G_{e}$
contains two $\mathbf{Z}_{2}$ generators, each of which determines a row of
$U$. Therefore, $U$ can also be completely determined in this case. The
corresponding mixing matrix can be found in \cite{toorop}.} As an example, let
$\{p_{1},p_{2},p_{3},m\}=\{3,5,2,5\}$ and $\{n_{1},n_{2},n_{3}%
,k\}=\{1,2,1,1\}$. From Eq.(\ref{twia}) and $\mathrm{Tr}[W_{i}]=-w_{i}$, one
has%
\[
w_{1}=0,\quad w_{2}=-1-2\cos[4\pi/5]=(\sqrt{5}-1)/2,\quad w_{3}=1.
\]
Then from Eq.(\ref{meq3}) and $4\sin^{2}\pi/5=10/(5+\sqrt{5})$, it follows
that%
\[
|\left(  U_{PMNS}\right)  _{\alpha1}|^{2}=\frac{5+\sqrt{5}}{10},\quad|\left(
U_{PMNS}\right)  _{\alpha2}|^{2}=\frac{5-\sqrt{5}}{10},\quad|\left(
U_{PMNS}\right)  _{\alpha3}|^{2}=0.
\]
Other elements can be obtained from Eq.(\ref{meq4}), and the resulting
$U_{PMNS}$ is a golden ratio mixing matrix (for explicit expression, see
\cite{toorop, a5m}). Note that one must set $\alpha=e=1$ for phenomenological
reasons. Also note that mixing matrices obtained by exchanging $p_{i}$ or
using a different value for $\alpha$ are just the original mixing matrix with
its rows and columns being reordered. But for the assignments discussed above
reordering is obviously phenomenologically unacceptable.

This example shows clearly the difference between the method discussed here
and the group theoretical method. The latter requires the knowledge of the
flavor group and its representation. Here what we need is a set of $p_{i}$ and
$n_{i}$ that satisfy the unitarity condition and no prior knowledge of the
flavor group is needed, which is to be determined by the solution. For
instance, the example discussed above corresponds to the solution
$\{p_{i}\}=\{2,3,5,5\}$ which indicates that the flavor group must contain
$\mathbf{Z}_{2}$, $\mathbf{Z}_{3}$, $\mathbf{Z}_{5}$, and $\mathbf{Z}%
_{2}\times \mathbf{Z}_{2}$ subgroups and its order must be a multiple of $60$.
It is not hard to find that the minimal finite group satisfying this condition
is $\mathbf{A}_{5}$. One can use $p_{i}$ and $n_{i}$ given above to verify
that this solution can indeed be realized by $\mathbf{A}_{5}$. Similarly, from
the solution leading to TBM, one can show that $\mathbf{S}_{4}$ is the minimal
horizontal symmetry that can produce TBM as a full-mixing matrix, which was
first pointed out in \cite{lam2}. In addition, one may also use $p_{i}$ and
$n_{i}$ to construct the group. In summary, the solution associated with a
mixing matrix can be regarded as the necessary conditions or minimal
requirements for a flavor symmetry to produce the mixing matrix. Although
necessary conditions are not sufficient to establish the existence of a finite
group, so far no counterexample has been found.

In the discussion above it is assumed that all $S_{i}$ and $T$ belong to
$G_{f}$, $\mathrm{Tr}[S_{i}T]$ and $\mathrm{Tr}[T]$ are real and the order of
$T$ is larger than $2$. As discussed in the previous section, although not
mandatory, they are general and reasonable assumptions on residual symmetries.
It is interesting to see that phenomenologically viable mixing patterns can
lead to include only several well-known ones. This result provides a different
way to understand those mixing patterns as natural consequences of discrete
flavor symmetries. On the other hand, the no-go result obtained above implies
that, to accommodate the mixing data including $\sin \theta_{13}$, which is
small but far from vanishing, some of the assumptions adopted in this section
may have to be relaxed, which will be discussed in the next section.

\section{Beyond the minimal scenario}

In this section, we focus on the consequences of allowing $\mathrm{Tr}%
[S_{i}T]$ or $\mathrm{Tr}[W_{i}]$ to be complex valued. In this case, the
eigenvalues of $W_{i}$ satisfying $\lambda^{p_{i}}=1$ can be written as
$e^{2\pi in_{i}/p_{i}}$, $e^{2\pi im_{i}/p_{i}}$ and $e^{-2\pi i(n_{i}%
+m_{i})/p_{i}}$. Without loss of generality, we require that $0\leq
n_{i},m_{i}<p_{i}$, $n_{i}+m_{i}\neq0$, and $\gcd(n_{i},p_{i})=1$. The
solutions to Eqs.(\ref{meq2}) and (\ref{meq1}) are given by%
\begin{align}
|\left(  U_{PMNS}\right)  _{\alpha i}|^{2}= &  \left(  \sin^{2}\frac{k\pi}%
{m}\right)  ^{-1}\cos \left[  \frac{n_{i}}{2p_{i}}2\pi \right]  \cos \left[
\frac{m_{i}}{2p_{i}}2\pi \right]  \cos \left[  \frac{n_{i}+m_{i}}{2p_{i}}%
2\pi \right]  ,\label{uai}\\
|\left(  U_{PMNS}\right)  _{\beta i}|^{2}= &  -\left(  2\sin^{2}\frac{k\pi}%
{m}\cos \frac{k\pi}{m}\right)  ^{-1}\cos \left[  \left(  \frac{k}{2m}%
+\frac{n_{i}}{2p_{i}}\right)  2\pi \right]  \nonumber \\
&  \cos \left[  \left(  \frac{k}{2m}+\frac{m_{i}}{2p_{i}}\right)  2\pi \right]
\cos \left[  \left(  \frac{k}{2m}-\frac{n_{i}+m_{i}}{2p_{i}}\right)
2\pi \right]  ,\label{ubi}\\
|\left(  U_{PMNS}\right)  _{\gamma i}|^{2}= &  -\left(  2\sin^{2}\frac{k\pi
}{m}\cos \frac{k\pi}{m}\right)  ^{-1}\cos \left[  \left(  \frac{k}{2m}%
-\frac{n_{i}}{2p_{i}}\right)  2\pi \right]  \nonumber \\
&  \cos \left[  \left(  \frac{k}{2m}-\frac{m_{i}}{2p_{i}}\right)  2\pi \right]
\cos \left[  \left(  \frac{k}{2m}+\frac{n_{i}+m_{i}}{2p_{i}}\right)
2\pi \right]  .\label{ugi}%
\end{align}
One can verify that Eqs.(\ref{meq3}) and (\ref{meq4}) are recovered when
$n_{i}$ or $m_{i}$ vanishes or $n_{i}+m_{i}=p_{i}$.

If all $S_{i}\in G_{f}$, as in the previous section, the unitarity condition
must be respected and then from Eq. (\ref{ucon0}), one has%
\begin{equation}
\sum \limits_{i=1}^{3}V_{i}+2\cos \alpha_{km}=-1,\quad V_{i}=\cos \alpha
_{n_{i}p_{i}}+\cos \alpha_{m_{i}p_{i}}+\cos \left(  \alpha_{n_{i}p_{i}}%
+\alpha_{m_{i}p_{i}}\right)  \label{uconi}%
\end{equation}
and%
\begin{equation}
\sum \limits_{i=1}^{3}V_{i}^{^{\prime}}=0,\quad V_{i}^{^{\prime}}=\sin
\alpha_{n_{i}p_{i}}+\sin \alpha_{m_{i}p_{i}}-\sin \left(  \alpha_{n_{i}p_{i}%
}+\alpha_{m_{i}p_{i}}\right)  \label{uconiz}%
\end{equation}
where $\alpha_{km}=2\pi k/m$, $\alpha_{n_{i}p_{i}}=2\pi n_{i}/p_{i}$, etc.
Unlike the minimal scenario where $\mathrm{Tr}[W_{i}]$ are real, the above two
equations are much more involved, and the complete set of solutions can hardly
be obtained in this case. But, as in the minimal scenario, solutions for
reasonably large $p_{i}$ can be exhausted by numerical calculations. Besides
that, they are also useful for one to approach possible solutions quickly,
especially when some $p_{i}$ are given.

As an example, let $m=4$ and $p_{1}=12$. Since $\cos(2\pi n_{1}/12)=\pm
\sqrt{3}/2$, $V_{1}$ and $V_{1}^{^{\prime}}$ are expected to be a quadratic
number\footnote{By quadratic number we mean a number satisfying a quadratic
equation with \emph{rational} coefficients.} of the form $a+b\sqrt{3}$ except
for particular choices of $n_{1}$ and $m_{1}$ [e.g., $n_{1}=1$ and $m_{1}=5$
or $6$, which lead to $|\left(  U_{PMNS}\right)  _{11}|=0$]. Therefore, to
satisfy Eq.(\ref{uconi}), it is reasonable to require that $p_{2}$ or $p_{3}$
also gives rise to some quadratic terms with the factor $\sqrt{3}$. The
simplest choice is to let, e.g. $p_{3}=12$. Then, without much effort, we find
that setting $p_{2}=3$ can lead to the mixing matrix
\[
\left \Vert U_{PMNS}\right \Vert =\frac{1}{2\sqrt{2}}\left(
\begin{array}
[c]{ccc}%
\sqrt{3+\sqrt{3}} & \sqrt{2} & \sqrt{3-\sqrt{3}}\\
\sqrt{2} & 2 & \sqrt{2}\\
\sqrt{3-\sqrt{3}} & \sqrt{2} & \sqrt{3+\sqrt{3}}%
\end{array}
\right)  .
\]
which presumably can be realized by the group $\mathbf{\Delta}(432)$.

Although the resulting $\sin^{2}\theta_{13}=0.158$ does not agree with the
data well, the point shown by this example is that if any $p_{i}$ grows large,
especially when the value of $\cos(2\pi n_{i}/p_{i})$ or $\sin(2\pi
n_{i}/p_{i})$ is not a rational number,\footnote{A more precise condition can
be given in terms of the algebraic degree of $\cos(2\pi n_{i}/p_{i})$ or
$\sin(2\pi n_{i}/p_{i})$, which is the lowest degree of the algebraic
equations with rational coefficients it satisfies. Except for $m=1$ or $2$,
the algebraic degree of $\cos(2\pi n/m)$ is given by $\varphi(m)/2$ \cite{wat}
where $\varphi(m)$ is Euler's $\varphi$-function or Euler's totient function
\cite{hardy}. For example, for $m=3$, $4$, $6$, $\varphi(m)/2=1$ and hence
$\cos(2\pi n/m)$ is a rational number of degree one. For $m=5$, $8$, $10$ and
$12$, $\varphi(m)/2=2$ and $\cos(2\pi n/m)$ is a quadratic number of degree
two. The degrees of numbers involved in algebraic operations, including
addition, product, division, etc. can provide valuable information about the
outcomes of these operations.} then for generic $n_{i}$ and $m_{i}$, it often
occurs that another $p_{i}$ has to acquire the same value. This also explains
to some extent why identical columns up to permutations often show up in
full-mixing matrices, as the one given above. Moreover, if $m$ takes a value
leading to a nonrational $\cos(2\pi k/m)$, e.g. $m=5$, $7$, etc. the above
unitary conditions are even harder to be satisfied since $m$ appears only in
one of them, i.e., Eq.(\ref{uconi}). This implies that even a large group can
lead to only a rather limited number of mixing patterns. A slight improvement
can be achieved by allowing $\mathrm{Tr}[T]$ to be complex, which will be
discussed later in this section.

As another interesting application, consider $m=3$ corresponding to
$G_{e}=\mathbf{Z}_{3}$ which occurs frequently. When $m=3$, one finds that
Eqs. (\ref{uai})--(\ref{ugi}) can be written as%
\begin{equation}
|\left(  U_{PMNS}\right)  _{\rho i}|^{2}=\frac{4}{3}\cos \left[  \frac{n_{\rho
i}}{2p_{\rho i}}2\pi \right]  \cos \left[  \frac{m_{\rho i}}{2p_{\rho i}}%
2\pi \right]  \cos \left[  \frac{n_{\rho i}+m_{\rho i}}{2p_{\rho i}}2\pi \right]
\label{uri}%
\end{equation}
where $\rho=\alpha$, $\beta$, or $\gamma$, $p_{\alpha i}=p_{i}$,$\quad
n_{\alpha i}=n_{i}$, $m_{\alpha i}=m_{i}$, and%
\begin{align*}
\frac{n_{\beta i}}{2p_{\beta i}} &  =\frac{1}{6}+\frac{n_{i}}{2p_{i}}%
,\quad \frac{m_{\beta i}}{2p_{\beta i}}=\frac{1}{6}+\frac{m_{i}}{2p_{i}},\\
\frac{n_{\gamma i}}{2p_{\gamma i}} &  =\frac{1}{6}-\frac{n_{i}}{2p_{i}%
}+H\left(  \frac{n_{i}}{2p_{i}}-\frac{1}{6}\right)  \\
\frac{m_{\gamma i}}{2p_{\gamma i}} &  =\frac{1}{6}-\frac{m_{i}}{2p_{i}%
}+H\left(  \frac{m_{i}}{2p_{i}}-\frac{1}{6}\right)
\end{align*}
in which $p_{\beta i}$ and $p_{\gamma i}$ are chosen to be the smallest
(positive) integers satisfying the above equations and $H(x)$ is the unit step
function defined as $H(x)=1$ (for $x>0$) and $H(x)=0$ (for $x<0$). As shown by
Eq.(\ref{uri}), the similarity among the expressions for $|\left(
U_{PMNS}\right)  _{\rho i}|^{2}$ indicates that $p_{\beta i}$ and $p_{\gamma
i}$ can also play the role of $p_{i}$. Therefore, by assigning $p_{i}$,
$p_{\beta i}$ and $p_{\gamma i}$ to the orders of $S_{i}T$, one can construct
a mixing matrix with identical columns and rows up to permutations. For
example, $\{p_{3},n_{3},m_{3},m,k\}=\{5,2,3,3,1\}$ results in $p_{\beta
i}=p_{\gamma i}=15$ and the squared mixing vector $\left.  \left(  6-2\sqrt
{5},\,3+\sqrt{5},\,3+\sqrt{5}\right)  \right/  12$. Then one may let
$p_{1,2}=15$ and the mixing matrix
\[
\left \Vert U_{PMNS}\right \Vert =\frac{1}{2\sqrt{6}}\left(
\begin{array}
[c]{ccc}%
\sqrt{5}+1 & \sqrt{5}+1 & \sqrt{10}-\sqrt{2}\\
\sqrt{10}-\sqrt{2} & \sqrt{5}+1 & \sqrt{5}+1\\
\sqrt{5}+1 & \sqrt{10}-\sqrt{2} & \sqrt{5}+1
\end{array}
\right)
\]
can be obtained by permuting the elements in the third column. A possible
issue of the mixing patterns being alike is that to fit the experimental data,
somewhat large corrections are required since in every column there is an
element equal to $|\left(  U_{PMNS}\right)  _{e3}|$. Also note that in this
example $\mathrm{Tr}[W_{3}]$ is real but $\mathrm{Tr}[W_{1,2}]$ are not. In
general, to produce mixing matrices like the one above, some $\mathrm{Tr}%
[W_{i}]$ must be complex valued; otherwise, the matrix $\left \Vert
U_{PMNS}\right \Vert $ should have two identical rows.

We now turn to phenomenological implications. As we know, most known
full-mixing patterns cannot fit the mixing data exactly within the
experimentally allowed range. Nevertheless, by using the method discussed
above, it is not hard to find one. For example, $\{p_{1},p_{2},p_{3}%
,m\}=\{20,3,10,3\}$ can lead to
\[
\left \Vert U\right \Vert =\frac{1}{2\sqrt{6}}\left(
\begin{array}
[c]{ccc}%
\sqrt{9+\sqrt{5}+c} & 2\sqrt{2} & \sqrt{7-\sqrt{5}-c}\\
\sqrt{6-2\sqrt{5}} & 2\sqrt{2} & \sqrt{10+2\sqrt{5}}\\
\sqrt{9+\sqrt{5}-c} & 2\sqrt{2} & \sqrt{7-\sqrt{5}+c}%
\end{array}
\right)
\]
where $c=\sqrt{6(5-\sqrt{5})}$. The mixing angles can be extracted as follows:
$\sin^{2}\theta_{13}=0.029$, $\sin^{2}\theta_{23}=0.62$, and $\sin^{2}%
\theta_{12}=0.34$. The obvious problem with this example is that it requires a
large group, presumably $\mathbf{\Delta}(600)$. In fact, this result is
confirmed by group theoretical calculation in a recent work \cite{Hol1} in
which a scan of groups with orders up to 1536 is performed. Our calculation
does not need any prior knowledge about $\mathbf{\Delta}(600)$ and, as
discussed above, it is the solution that suggests $\mathbf{\Delta}(600)$ as a
candidate. Besides $\mathbf{\Delta}(600)$, the other two mixing patterns found
in \cite{Hol1} for $\mathbf{\Delta}(1536)$ and $(\mathbf{Z}_{18}%
\times \mathbf{Z}_{6})\rtimes \mathbf{S}_{3}$ can also be obtained from
Eqs.~(\ref{uai})--(\ref{ugi}). Just in case it is needed, below we give the
exact value of the mixing matrix for $\mathbf{\Delta}(1536)$%
\[
\left \Vert U\right \Vert =\frac{1}{\sqrt{6}}\left(
\begin{array}
[c]{ccc}%
\sqrt{2+\sqrt{2+\sqrt{2}}} & \sqrt{2} & \sqrt{2-\sqrt{2+\sqrt{2}}}\\
\sqrt{2-c_{-}} & \sqrt{2} & \sqrt{2+c_{-}}\\
\sqrt{2-c_{+}} & \sqrt{2} & \sqrt{2+c_{+}}%
\end{array}
\right)
\]
where $c_{\pm}=\sqrt{2\pm \sqrt{2\mp \sqrt{3}}}$. The numerical values of
$\sin^{2}\theta_{ij}$ can be found in \cite{Hol1}. As to the group
$(\mathbf{Z}_{18}\times \mathbf{Z}_{6})\rtimes \mathbf{S}_{3}$, the
corresponding mixing matrix does not admit an expression in terms of exact
values as the one shown above\footnote{Using algebraic number theory, one can
show that mixing elements can be expressed in terms of exact values only when
some conditions are satisfied. For example, in the case where $m=3$, the
orders of $S_{i}T$ must be products of Fermat numbers. By exact values we mean
numbers that can be expressed in terms of rational numbers involving only
sums, products and square roots, such as $(2-\sqrt{3})/3$. This also implies
that among all the mixing patterns that can be expressed in exact forms, only
very particular ones such as those discussed in this paper, can be produced by
discrete symmetries. Therefore, one can show that some mixing patterns, e.g.
the hexagonal mixing, cannot be produced by discrete symmetries at least in
the framework discussed in this work. More detailed discussion will be given
elsewhere.}.

Besides \cite{Hol1}, it is also reported in \cite{Lam1} that no $SU(3)$
subgroup of order less than $512$ can produce the full-mixing data. For us
this is almost an expected result since from Eqs.(\ref{uai})--(\ref{ugi}) it
is easy to see that small $\sin \theta_{13}$ requires one or more angles
involved to be close to $\pi/2$ or $3\pi/2$, and hence, together with the
reality conditions, it would require $p_{i}$ or $m$ to be large. Equations
(\ref{uai})--(\ref{ugi}) can also be evaluated numerically and the result
agrees with them, as expected. Since in general, mixing parameters also
receive contributions from higher-order corrections including radiative
corrections, this result implies that it might be more plausible to introduce
sizable corrections (or free parameters) to the mixing patterns obtained from
discrete symmetries, which can be considered as leading order contributions
\cite{xing1}. Partial mixing is also a reasonable option, as\ discussed in
\cite{Sim1, Sim2}.

Finally, we comment on the case where $\operatorname{Im}(\mathrm{Tr}[T])\neq
0$. When it occurs, the unitarity conditions given by Eqs.(\ref{uconi}) and
(\ref{uconiz}), which follow from Eq.(\ref{ucon0}), should be modified to%
\begin{equation}
\sum \limits_{i=1}^{4}V_{i}=0,\quad V_{i}=\cos \alpha_{n_{i}p_{i}}+\cos
\alpha_{m_{i}p_{i}}+\cos \left(  \alpha_{n_{i}p_{i}}+\alpha_{m_{i}p_{i}%
}\right)  \label{uconit}%
\end{equation}
and%
\begin{equation}
\sum \limits_{i=1}^{3}V_{i}^{^{\prime}}=V_{4}^{^{\prime}},\quad V_{i}%
^{^{\prime}}=\sin \alpha_{n_{i}p_{i}}+\sin \alpha_{m_{i}p_{i}}-\sin \left(
\alpha_{n_{i}p_{i}}+\alpha_{m_{i}p_{i}}\right)  .\label{uconitz}%
\end{equation}
Although more complicated and even harder to be satisfied, it allows new
mixing patterns for particular $p_{i}$. As an example, let $m=p_{4}=7$,
$n_{4}=1$. The quickest way to find a solution in this case is to use the
trick of assigning the value of $p_{4}$ or $m$ to another $p_{i}$, which
fortunately works again. After that, the remaining $p_{i}$ can be easily
derived. One can verify that a solution can be found and it leads to another
$PSL(2,Z_{7})$ mixing matrix given in \cite{toorop} [see Eq.~(40) there].

Although a solution is found in the example above, it is not hard to see that
it is very likely that this solution is the only solution for the case where
$m=7$, which shows that if $m$ is associated with a $\cos(2\pi/m)$ with a high
degree (see footnote 6), the unitary conditions are hard to be satisfied and
solutions may not always exist. It also provides further support to our
finding that using large groups may not help too much for the purpose of
producing full-mixing matrices that can fit all the experimental data. One may
wonder whether good chances would appear if the restriction to $SU(3)$ is
removed. However, it does not seem to be very likely because if $\det T\neq1$,
then $p_{i}$ are constrained not only by the unitarity condition which will
become more complicated, but also by the relation Eq.(\ref{ba}). Since this is
the case beyond the scope of this paper, we will leave it for future work.

\section{Conclusions}

As shown in the previous sections, the method discussed in this paper can
provide an efficient way to derive neutrino mixing patterns determined by
underlying discrete flavor symmetries. Although most of our attention is paid
to full-mixing patterns, the formalism developed in this paper can also be
used to calculate mixing elements in the partial-mixing case. It can be used
to cross-check with group theoretical calculations or as a simpler way to
reach some known results. It also gives the necessary conditions that can lead
to the flavor symmetry corresponding to a mixing matrix or mixing-matrix
column derived by this method. Since the properties of most $SU(3)$ subgroups
can be found in existing literatures, e.g. \cite{grimus2, ish, ludl}, one may
find an appropriate group without much trouble in most cases and the physical
viability can be verified concretely, as shown by the $\mathbf{\Delta}(600)$ example.

Moreover, it can also shed new light on the relation between neutrino mixing
and discrete symmetries. We find that in the minimal scenario where some
general assumptions are adopted, except phenomenologically disfavored mixing
patterns, discrete symmetries cannot lead to mixing patterns other than the
well-known TBM, BM and golden ratio mixings. This result not only provides a
new way to understand these mixing patterns as simple and natural consequences
of discrete symmetries, but also clearly indicates the limitation of the
minimal scenario and how to go beyond it. Another interesting result can be
found by this method is that, although relaxing some assumptions can lead to
new mixing patterns, to fit the mixing data, relative large groups are
required for full-mixing matrices unless sizable corrections are allowed.

\section*{Acknowledgements}

This work was supported in part by the National Science Foundation of China
(NSFC) under the Grant No. 10965003.

\end{document}